\documentclass[11pt]{article}
\usepackage{moriond,epsfig}
\def\be{\begin{equation}}
\def\ee{\end{equation}}
\def\bc{\begin{center}}
\def\ec{\end{center}}
\def\bea{\begin{eqnarray}}
\def\eea{\end{eqnarray}}

\def\nn{\nonumber}
\def\ov{\overline}

\def\drb{$\overline{\rm DR}$}

\def\sq2{\sqrt{2}}

\def\smlR{\scriptscriptstyle R}
\def\smlDR{\scriptscriptstyle \overline{DR}}

\def\li2{{\rm Li}_2}

\def\EP{effective potential}

\def\as{\alpha_s}
\def\at{\alpha_t}

\def\og2{{\cal O}(g^2)}
\def\oat{{\cal O}(\at)}
\def\oas{{\cal O}(\as)}

\def\oats{{\cal O}(\at \as)}
\def\oatq{{\cal O}(\at^2)}
\def\oatstq{{\cal O}(\at \as +\at^2 )}

\def\mgl{m_{\tilde{g}}}
\def\msqu{m_{\tilde{t}_1}^2}
\def\msqd{m_{\tilde{t}_2}^2}
\def\msqi{m_{\tilde{t}_i}^2}

\def\diff{\msqu-\msqd}

\def\mtu{m_{\tilde{t}_1}}
\def\mtd{m_{\tilde{t}_2}}
\def\mz{m_{\scriptscriptstyle Z}}
\def\ma{m_{\scriptscriptstyle A}}

\newcommand{\mabar}{\overline{m}_{\scriptscriptstyle A}}
\newcommand{\smallz}{{\scriptscriptstyle Z}}
\def\mh{m_h}
\def\mt{m_t}

\def\A0{\ma^2}

\def\thz{\theta_{\tilde{t}}}

\def\s2t{s_{2\theta}}
\def\c2t{c_{2\theta}}
\def\ss2t{s_{2\bar{\theta}}}
\def\cc2t{c_{2\bar{\theta}}}

\def\sb{s_\beta}
\def\cb{c_\beta}

\def\logA0{\ln\frac{\,\A0}{Q^2}}

\def\logqA0{\ln^2\frac{\A0}{Q^2}}

\begin{document}
\vspace*{0.5cm}
\title{TWO-LOOP CORRECTIONS TO THE HIGGS \\ BOSON MASSES IN THE MSSM}

\author{ P. SLAVICH }

\address{Physikalisches Institut der Universitat Bonn,\\
Nussallee 12, D-53115 Bonn, Germany}

\maketitle\abstracts{We present a computation of the $\oatstq$
two--loop corrections to the MSSM Higgs masses. An appropriate use of
the effective potential approach allows us to obtain simple analytical
formulae, valid for arbitrary values of $\ma$ and of the mass
parameters in the stop sector. In a large region of the parameter
space the $\oatq$ corrections are comparable to the $\oats$ ones,
increasing the prediction for $\mh$ by several GeV.}

\vspace*{1cm}
\begin{center}
{\em Talk given at the XXXVII Rencontres de Moriond,\\ 
Electroweak Interactions and Unified Theories, \\
Les Arcs, France, March 9--16 2002.}
\end{center}
\vspace*{1.5cm}

In the Minimal Supersymmetric Standard Model (MSSM), the Higgs sector
contains two $SU(2)_L$ doublets, and the masses and the couplings of
the Higgs bosons can be expressed, at the classical level, in terms of
known Standard Model quantities and only two additional parameters:
for example, the mass of the neutral CP--odd state, $\ma$, and the
ratio of the two Higgs vacuum expectation values, $\tan \beta$.
In particular, the mass $\mh$ of the lightest Higgs boson is predicted
to be smaller than the mass of the $Z$ boson. However, radiative
corrections modify considerably the predictions for the Higgs masses,
allowing the light Higgs mass to escape the tree--level bound, and
bring along additional dependences on the remaining MSSM
parameters. Direct searches at LEP have already ruled out a
considerable fraction of the MSSM parameter space, and the forthcoming
high--energy experiments at the Tevatron and the LHC will either
discover (at least) one light Higgs boson or rule out the MSSM as a
viable theory for physics at the weak scale. Thus, a precise
calculation of the radiative corrections to the Higgs boson masses,
valid for arbitrary values of all the relevant parameters, is needed
to compare reliably the MSSM predictions with the (present and future)
experimental data.

In this talk, a computation \cite{DSZ,BDSZ} of both $\oats$ and
$\oatq$ two--loop corrections to the MSSM Higgs boson masses, based on
the \EP\ approach, is presented. The results of refs.~\cite{DSZ,BDSZ}
extend the existing two--loop calculations
\cite{hh}\mbox{}$^{-}$\cite{ez2} to the case of arbitrary MSSM
parameters, and are given as a set of simple analytical formulae,
suitable for being promptly included in the experimental analyses.

We start our discussion of the corrections to the Higgs masses by
recalling that in the \EP\ approach the CP--odd and CP--even mass
matrices are identified with the second derivatives of the effective
potential, $V_{\rm eff} = V_0 + \Delta V$, evaluated at its minimum: \be
\label{defmat}
\left({\cal M}^2_P\right)_{ij}^{\rm eff} \equiv
\left. \frac{\partial^2  V_{\rm eff}}{\partial P_i \partial P_j}
\right|_{\rm min} \, , 
\hspace{1cm}
\left({\cal M}^2_S\right)_{ij}^{\rm eff} \equiv
\left. \frac{\partial^2  V_{\rm eff}}{\partial S_i \partial S_j}
\right|_{\rm min} \, ,
\hspace{1cm}
(i,j=1,2) \, ,
\ee
where we have decomposed the Higgs fields into their VEVs plus their 
CP--even and CP--odd fluctuations as 
$H_i^0 \equiv (v_i + S_i + i \, P_i)/\sq2\,$.
Using the explicit expression of the tree--level potential $V_0$,
and imposing minimization conditions on $V_{\rm eff}$, the mass matrix
for the CP--even sector can be written as:

\be
\left( {\cal M}^2_S \right)^{\rm eff}  = 
\left( {\cal M}^2_S \right)^{0, \, {\rm eff}} 
+
\left(\Delta{\cal M}^2_S\right)^{\rm eff}  \, ,
\label{effmatrix}
\ee
where
\be
\left( {\cal M}^2_S \right)^{0, \, {\rm eff}}
= 
\left(\begin{array}{cc}
\ov m_\smallz^2 \, \cb^2 + \ov \ma^2 \,\sb^2 
& -\left(\ov m_\smallz^2 + \ov \ma^2 \right) \sb\, \cb \\ 
-\left(\ov m_\smallz^2 + \ov \ma^2\right) \sb\, \cb 
& \ov m_\smallz^2 \, \sb^2 + \ov \ma^2 \,\cb^2 
\end{array}\right) 
\label{mpma} 
\, ,
\ee

\be
\left(\Delta{\cal M}^2_S\right)_{ij}^{\rm eff}  = 
\left.\frac{\partial^2 \Delta V}{\partial S_i \partial S_j}
\right|_{\rm min} -(-1)^{i+j}\left.
\frac{\partial^2 \Delta V}{\partial P_i \partial P_j}\right|_{\rm min} \, .
\label{Dms}
\ee

In Eq.~(\ref{mpma}), $\mabar^2$ is the mass of the CP--odd state in
the \EP\ approximation, i.e. the non--vanishing eigenvalue of $\left(
{\cal M}^2_P \right)^{\rm eff}$. Similarly, $\ov{m}_\smallz^2 = (g^2 +
g'^2) \, v^2/4$ is just the $\ov{\rm DR}$ mass for the $Z$
boson. Finally, $\sb \equiv \sin\beta$ and $\cb \equiv \cos\beta$,
where $\tan\beta = v_2/v_1$.

According to Eq.~(\ref{Dms}), $\left( \Delta{\cal M}^2_S \right)^{\rm
eff}$ can be computed by taking the derivatives of $\Delta V$ with
respect to the CP--even and CP--odd fields, evaluated at the minimum
of $V_{\rm eff}$. In this computation we express the field--dependent
masses and interaction vertices that contribute to $\Delta V$, and are
relevant for our calculation, in terms of five field--dependent
quantities. The first three, appearing already in the one-loop
contribution to the \EP, are the squared masses of the top quark and
of the stop squarks. The remaining two quantities, appearing only in the
two-loop contribution to the \EP, can be chosen as the stop mixing
angle, denoted as $\thz$, and $\cos\,(\varphi - \tilde{\varphi})$,
where $\varphi$ is the phase in the complex top mass, and
$\widetilde{\varphi}$ is the phase in the off--diagonal term of the
stop mass matrix.

After a straightforward although lengthy application of the
chain rule for the derivatives of the effective potential, we derive
an expression for the two--loop corrections to the Higgs mass matrix,
valid when the one--loop part of the corrections is written in terms
of one--loop renormalized parameters:

\be
\label{dmsren}
\left(\Delta {\cal M}^2_S\right)^{\rm eff} =  
\sqrt{Z} \, \left( \Delta {\cal \widehat M}^2_S 
\right)^{\rm eff}  \sqrt{Z} \, ,
\ee
where $Z$ is a wave function renormalization matrix for the Higgs fields,
and
\bea
\label{dms11ren}
\hspace{-1cm}\left(\Delta {\cal \widehat M}^2_S\right)_{11}^{\rm eff} 
\!\!& = &\!\!
\frac{1}{2} \, h_t^2 \,\mu^2 \,\s2t^2  \,
\left(F_3 + 2 \, \Delta_\mu F_3 \right)  \, , \\
&&\nn\\
\label{dms12ren}
\hspace{-1cm}\left(\Delta {\cal \widehat M}^2_S\right)_{12}^{\rm eff} 
\!\!& = &\!\!
h_t^2 \,\mu\, m_t\, \s2t \,  \left( F_2 +  \Delta_\mu F_2 \right)
+ \frac{1}{2}\, h_t^2\, A_t \,\mu \, \s2t^2 \, \left(F_3 
+ \Delta_{A_t} F_3 + \Delta_\mu F_3 \right) \, , \\
&&\nn\\
\label{dms22ren}
\hspace{-1cm}\left(\Delta {\cal \widehat M}^2_S\right)_{22}^{\rm eff} 
\!\!& = &\!\!
2\, h_t^2\, m_t^2\, F_1
+ 2\, h_t^2\, A_t\, m_t\, \s2t\, 
\left( F_2 + \Delta_{A_t} F_2 \right)
+ \frac{1}{2}\, h_t^2\, A_t^2\, \s2t^2\, 
\left(F_3 + 2\, \Delta_{A_t} F_3 \right).
\eea
In the above equations the functions $(F_1,F_2,F_3)$ are written as
$F_i = \widetilde{F_i} + \Delta \widetilde{F_i}$, where
$\widetilde{F_i}$ are combinations of the derivatives of the pure
two--loop effective potential whose explicit definitions are given in
ref.~\cite{DSZ}. The terms $\Delta \widetilde{F_i}$ include
contributions coming from the renormalization of the top and stop
masses appearing in the one--loop part of the $F_i$, as well as from
the renormalization of the common factors multipliying $F_i$,
i.e. ($h_t^2\,, m_t\,,\s2t$).  On the other hand, in each entry of
$\left(\Delta {\cal \widehat M}^2_S\right)^{\rm eff}$ the functions
$F_i$ are multiplied by different combinations of $\mu$ and
$A_t$. These two parameters have different $\oat$ and $\oas$
renormalizations that cannot be absorbed in the $F_i$, and are then
separately taken into account by the terms $\Delta_\mu F_i$ and
$\Delta_{A_t} F_i$.

We have derived explicit analytical formulae
for the $\oats$ and $\oatq$ parts of the functions $F_i$, obtained
under the assumption that the $\oat$ parts are expressed in terms of
\drb\ quantities.  In ref.~\cite{DSZ} we presented the complete formulae
for the $\oats$ corrections, valid for arbitrary choices of all the
input parameters. On the other hand, the general formulae for the
$\oatq$ corrections are quite long, thus we made them available upon
request in the form of a computer code. In ref.~\cite{BDSZ}, we
presented explicit analytical formulae valid for $m_Q = m_U \equiv
M_S$, in which ${\cal O}(m_t/M_S)$ corrections are neglected,
but $\ma$ is left arbitrary.

In the phenomenological analyses of the MSSM Higgs sector, it may be
useful to express or results in terms of input parameters given in a
different renormalization scheme, which will be indicated generically
as R. To obtain the ${\cal O}(\at \as + \at^2)$ correction in the R
scheme one has just to shift the parameters appearing in the
one--loop term, i.e. $x^{\smlDR} = x^{\smlR} + \delta x$, where $x$ is
a generic parameter. This induces additional two--loop contributions
that can be absorbed in a redefinition of the functions $F_i$ and
$\Delta F_i$ entering Eqs.~(\ref{dms11ren})--(\ref{dms22ren}).
Explicit formulae for the transition from the \drb\ scheme to a
generic R scheme are given in refs. \cite{DSZ,BDSZ}.

The parameters entering the one--loop part of the corrections can be
identified as ($\mt$, $\mtu$, $\mtd$, $\thz$, $A_t$, $\mu$, $v$,
$\tan\beta$). Some of them are related by the equality 
\be
\label{sinmin}
\sin 2 \thz = \frac{2\,m_t\,(A_t + \mu\cot\beta)}{\diff}\;.
\ee
We choose to express our result identifying the stop
masses $\msqi$ and the top mass $\mt$ with the corresponding pole
masses. For the electroweak symmetry--breaking parameter $v$, we use
the value obtained in terms of the precisely known muon decay constant
$G_{\mu}$.  These choices specify $\delta \msqi$, $\delta \mt$ and
$\delta v$ in terms of the finite parts of stop, top and W-boson
self-energies respectively. For the stop mixing angle, several OS
prescriptions are present in the literature. We find
suitable for our $\oatstq$ calculation the following `symmetrical'
definition \cite{mixing}
\be
\label{thyam}
\delta\thz =  \frac{1}{2}\, 
\frac{{\Pi}_{12}(\msqu) + {\Pi}_{12}(\msqd)}{\msqu-\msqd} \, ,
\label{detheta}
\ee
where ${\Pi}_{12}(p^2)$ stands for the finite part of the
off--diagonal stop self--energy.

It is not clear what meaning should be assigned to an OS definition of
the remaining quantities, i.e. $(A_t,\mu,\tan\beta)$.  For instance,
they could be related to specific physical amplitudes. However, given
our present ignorance of any supersymmetric effect, such a choice does
not seem particularly useful.  Then, we find simpler to assign the
quantities $\mu$ and $\tan\beta$ in the \drb\ scheme, at a reference
scale $Q_0 = 175$ GeV, that we choose near the present central value of
the top quark mass.  In this framework, we are going to take $\delta
\mu = \delta \beta = 0$, and we treat $A_t$ as a derived quantity,
obtained through Eq.~(\ref{sinmin}). 

Our effective potential calculation is equivalent to the evaluation of
the Higgs self--energies in the limit of vanishing external
momentum. A diagrammatic computation of the two--loop $\oats$
contributions to the Higgs boson self--energies at zero external
momentum has been performed in ref. \cite{hollik}. Analytical formulae,
valid in the simplified case of degenerate soft stop masses and zero
mixing (with $\mu = A_t = 0$), have been presented in the first paper
of ref.~\cite{hollik}. For arbitrary values of the top and stop
parameters, however, the complete analytical result of ref. \cite{hollik}
is far too long to be explicitly presented, and is only available as a
computer code \cite{feynhiggs}. We have checked that, in the case of
zero mixing and degenerate stop masses, our results coincide with
those of ref. \cite{hollik}. Moreover, after taking into account the
difference in the definitions of the OS renormalized angle $\thz$, we
find perfect agreement with the numerical results of ref. \cite{feynhiggs},
for arbitrary values of all the input parameters.

A calculation of both $\oats$ and $\oatq$ two--loop corrections to the
lightest Higgs boson mass $m_h$, based on the formalism of the
effective potential, has been presented in refs. \cite{ez1,ez2}.  In
these papers, however, the dependence of the stop masses and mixing
angles on the fields $P_i$ (the CP--odd components of the neutral
Higgs fields) is not taken into consideration. Therefore, the
$\oatstq$ corrections to the parameter $m_A$ are not evaluated.  If
one wants to relate the input parameters to measurable quantities, the
computation is applicable only in the limit $m_A \gg m_Z$, in which
$m_h$ is nearly independent of $m_A$. Moreover, the results of
refs. \cite{ez1,ez2} for $m_h$ are available in numerical form, and
simple analytical formulae are provided only in the limit of universal
soft stop masses (degenerate with $m_A$ and $\mgl$) much larger than
the top quark mass. We have verified that, in such limit, our
analytical results for $\mh$ agree with those of refs. \cite{ez1,ez2}.

\begin{figure}[t]
\begin{center}
\mbox{
\hspace{-.4cm}
\epsfig{figure=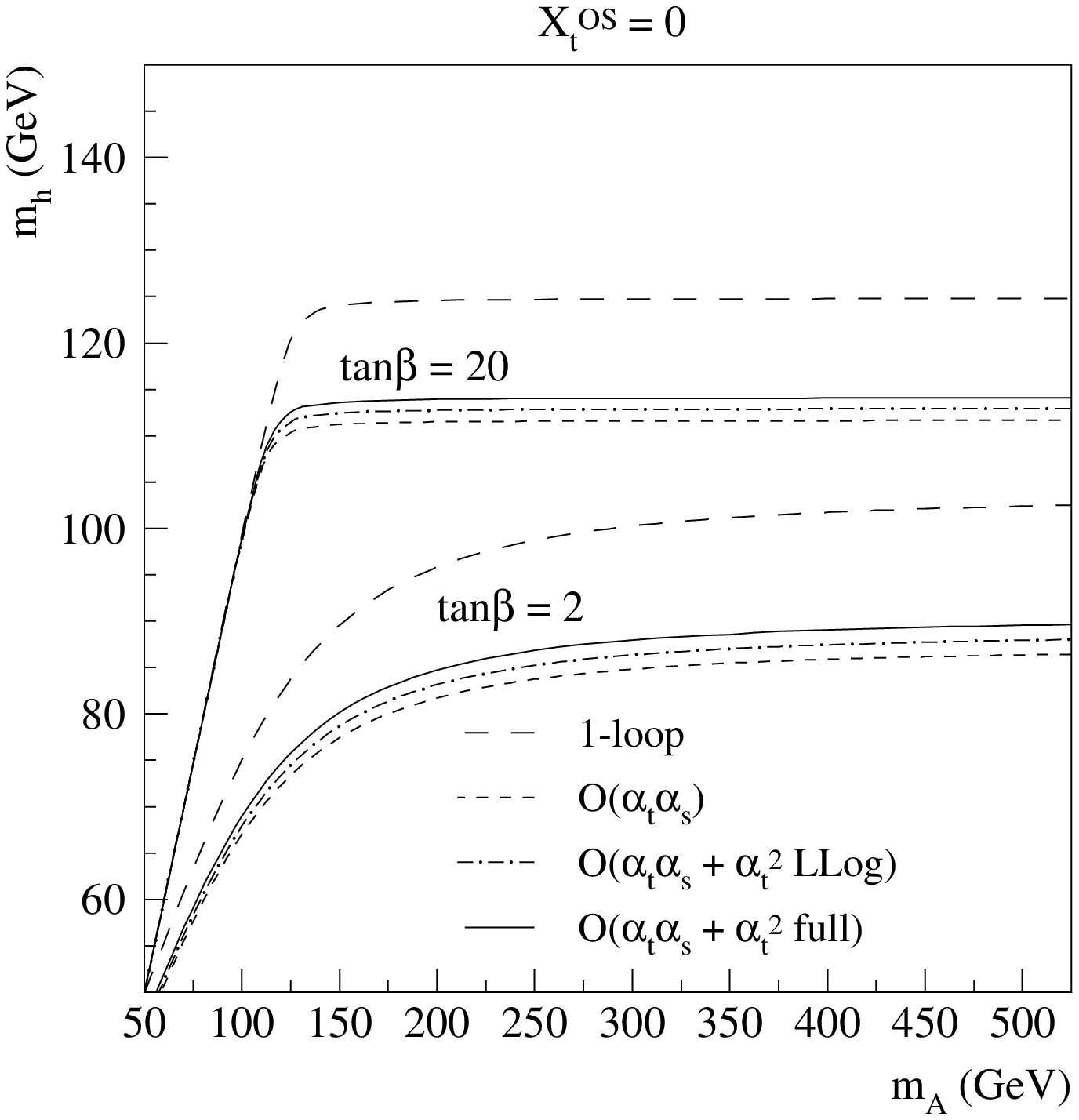,width=9cm,height=9cm}
\hspace{-1cm}
\epsfig{figure=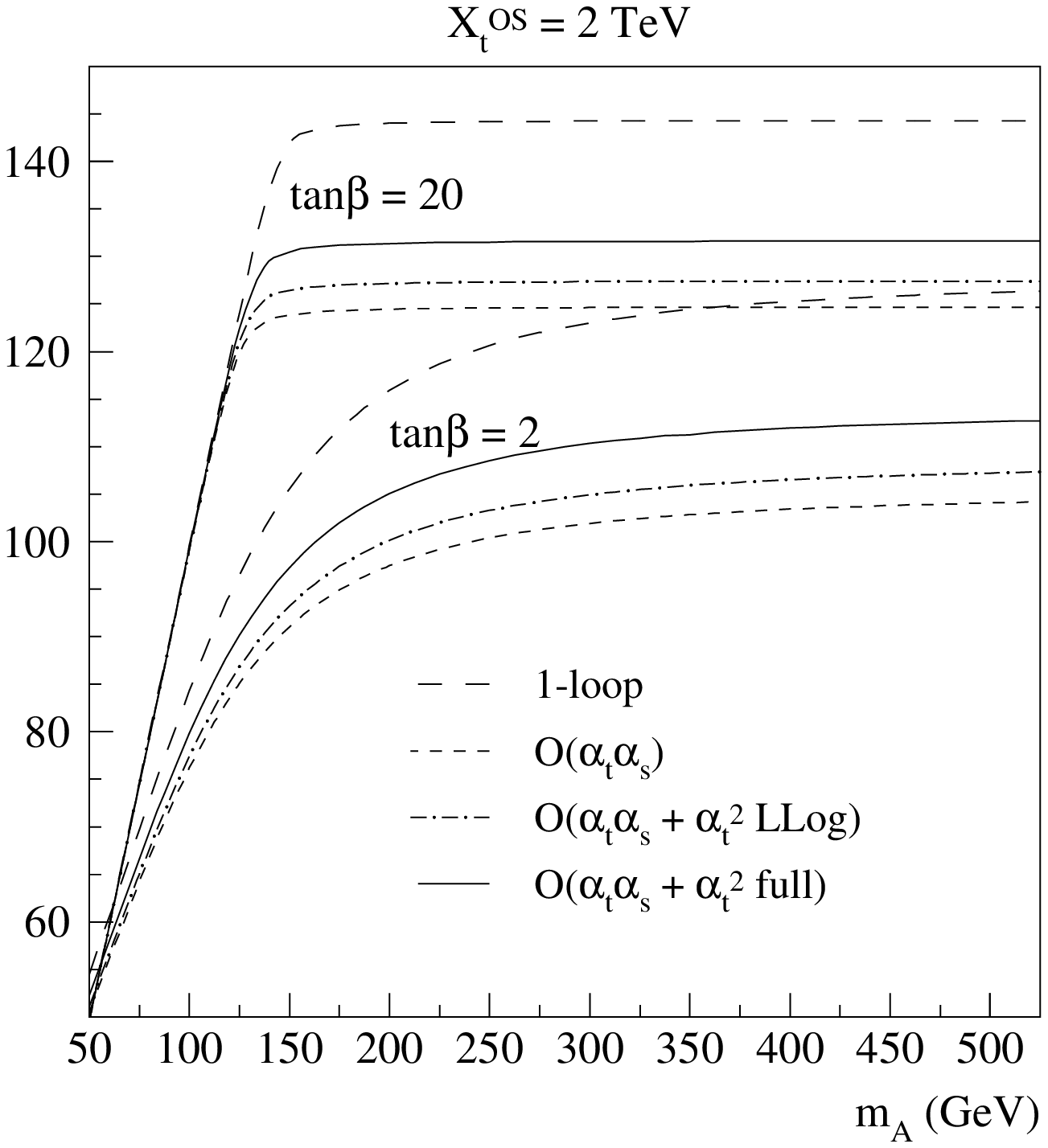,width=9cm,height=9cm}}
\end{center}
\vspace{-0.5cm}
\caption{The mass $m_h$ as a function of $\ma$,
for $\tan\beta = 2$ or 20 and $X_t^{\rm OS} = 0$ or 2 TeV.
The other parameters are $m_Q^{\rm OS} = m_U^{\rm OS} = 1$ TeV, 
$\mu = 200$ GeV, $\mgl = 800$ GeV. The meaning of the 
curves is explained in the text.}
\label{mhvsma}
\end{figure}

In the analysis of ref. \cite{hollik} some `leading logarithmic' ${\cal
O}(\at^2)$ corrections, obtained by renormalization group methods 
\cite{rge}, have indeed been added to $\left({\cal M}_S^2\right)_{22}$.  
However, when comparing our complete ${\cal O}(\at^2)$ result with the 
renormalization group result, it turns out
that, for some choices of the SUSY parameters, the leading logarithmic
corrections amount only to a fraction of the full ones.  
Fig.~\ref{mhvsma} shows $m_h$ as a function of $\ma$, for $\tan \beta
= 2$ or 20 and \footnote{To facilitate the comparison with the existing
analyses of the two--loop corrected Higgs masses, we make use of the
unphysical parameters $(m_Q^{\rm OS}, m_U^{\rm OS}, X_t^{\rm OS})$
that can be derived by rotating the diagonal matrix of the OS stop
masses by an angle $\thz$.} $X_t^{\rm OS} = 0$ or 2 TeV. The other
input parameters are chosen as $m_t = 175$ GeV, $m_Q^{\rm OS} =
m_U^{\rm OS} = 1$ TeV, $\mu = 200$ GeV, $\mgl = 800$ GeV. The cases
with $X_t^{\rm OS} = 0$ and $X_t^{\rm OS} = 2$ TeV correspond,
respectively, to the so-called `no--mixing' and `$m_h$--max' benchmark
scenarios \cite{bench} considered in the experimental analyses
\cite{lepexp}. The curves in Fig.~\ref{mhvsma} correspond to the
two--loop corrected Higgs mass at ${\cal O}(\at\as)$ (short--dashed
line), at ${\cal O}(\at\as+\at^2)$ including only the leading
logarithmic corrections (dot--dashed line), and at ${\cal
O}(\at\as+\at^2)$ including our full computation (solid line).  The
one--loop result for $m_h$ is also shown for comparison (long--dashed
line). It can be seen from the figure that the ${\cal O}(\at\as)$
corrections are in general large, reducing the one--loop result for
$m_h$ by 10--20 GeV. On the other hand, the ${\cal O}(\at^2)$
corrections tend to increase $m_h$: for small stop mixing they are
generally small (less than 2--3 GeV), whereas for large stop mixing
they can reach 7--8 GeV, i.e. a non--negligible fraction of the ${\cal
O}(\at\as)$ ones.  Moreover, it appears that the `non--leading'
corrections included in our full result are always comparable in size
with the `leading logarithmic' ones, and are even more significant
than the latter in the case of large stop mixing. For $\ma \gg \mz$,
we find good numerical agreement with the value of $m_h$ obtained from
the approximate formulae of ref.~\cite{ez2}.

In conclusion, refs.~\cite{DSZ,BDSZ} contain the most advanced
calculation (so far) of the two--loop corrections to the MSSM neutral
Higgs boson masses. Our work should lead to a more accurate interpretation
of the experimental searches for the neutral MSSM Higgs bosons at LEP,
the Tevatron, the LHC and other possible future colliders. The
importance of the new ${\cal O}(\at^2)$ effects we have computed will
increase further when the top quark mass is measured more precisely.
Then, we can hope for the next step, the discovery of supersymmetric
particles and supersymmetric Higgs bosons.

\section*{Acknowledgments}
The author thanks the organizers of the XXXVII Rencontres de Moriond
for the pleasant atmosphere in which the talk was given. This work was
supported in part by the European Union under the contracts
HPRN-CT-2000-00149 (Collider Physics) and HPRN-CT-2000-00148 (Across
the Energy Frontier).

\end{document}